\documentclass{article}

\usepackage{PRIMEarxiv}

\usepackage[utf8]{inputenc} 
\usepackage[T1]{fontenc}    
\usepackage{hyperref}       
\usepackage{url}            
\usepackage{booktabs}       
\usepackage{amsfonts}       
\usepackage{nicefrac}       
\usepackage{microtype}      
\usepackage{lipsum}
\usepackage{fancyhdr}       
\usepackage{graphicx}       
\usepackage{xcolor}         
\usepackage{color}
\usepackage{wrapfig}
\usepackage{amsmath,amssymb}
\usepackage{subfigure}
\usepackage{makecell}
\usepackage{enumerate}

\title{Adaptive Retrieval Strategies for Biomedical Question Answering}

\author{%
    Han Yue
    \And
    Eric Zhou
    \And
    Alex Hu
    \And
    Xueshen Li
    \And
    Yuan Zhang
    \And
    Jinfeng Zhang
    \And
    \texttt{\{hyue,eric,alexhu,xli,yuan,jinfeng\}@insilicom.com}
}

\begin{document}
\maketitle

    \begin{abstract}
    Biomedical question answering (QA) encompasses diverse question types, including yes/no, factoid, list, and summary questions, each requiring distinct forms of evidence and reasoning. However, most retrieval-augmented QA systems rely on a unified retrieval pipeline, regardless of the information needs of different question categories. This one-size-fits-all approach may limit the effectiveness of evidence acquisition and downstream answer generation. In this work, we propose an adaptive retrieval framework that selects retrieval and evidence aggregation strategies according to question type. The system combines query understanding, biomedical document retrieval, reranking, knowledge graph augmentation, document clustering, and large language model-based answer generation. For yes/no questions, it focuses on precise evidence retrieval; for factoid and list questions, it emphasizes entity-oriented retrieval and clustering; and for summary questions, it performs broader evidence collection and synthesis. We evaluate the proposed framework on the BioASQ benchmark and demonstrate that adaptive retrieval strategies improve evidence relevance and answer quality across multiple question types. Our results suggest that aligning retrieval mechanisms with question-specific information needs provides an effective direction for enhancing retrieval-augmented biomedical QA systems.
    \end{abstract}

\section{Introduction}

    The BioASQ Challenge\footnote{\href{https://www.bioasq.org/}{https://www.bioasq.org/}} is an annual competition focused on advancing biomedical question answering and semantic indexing systems. Over the years, the challenge has continuously evolved to include new features and refine its evaluation metrics:

    \begin{itemize}
        \item Since BioASQ 8, Phase A evaluation uses Mean Average Precision (MAP), limiting results to 10 elements per question.
        \item Since BioASQ 9, snippet evaluation in Phase A is based on the mean F-measure.
        \item Starting from BioASQ 12, a new Phase A+ was introduced, allowing participants to submit answers before gold documents and snippets are available.
    \end{itemize}
    
    The BioASQ 13B Challenge\footnote{\href{https://participants-area.bioasq.org/general_information/Task13b/}{https://participants-area.bioasq.org/general\_information/Task13b/}} builds on these improvements and uses benchmark datasets curated by biomedical experts, aiming to push the boundaries of biomedical information retrieval and question answering. It consists of three main phases: Phase A, Phase A+, and Phase B. Details are as follows.

    \textbf{Phase A}. In Phase A of Task 13b, the participants will be provided with English questions $q_1$, $q_2$,...,$q_n$. For each question $q_i$, each participating system will be required to return any (ideally all) of the following lists:
    \begin{itemize}
        \item A list of at most 10 relevant articles (documents) $d_{i,1}$, $d_{i,2}$, $d_{i,3}$,... from the designated article repositories. Again, the list should be ordered by decreasing confidence, i.e., $d_{i,1}$ should be the article that the system considers to be the most relevant to the question $q_{1}$, $d_{i,2}$ should be the article that the system considers to be the second most relevant etc. A single article list will be returned per question and participating system, and the list may contain articles from multiple designated repositories. The returned article list will actually contain unique article identifiers (obtained from the repositories).
        \item A list of at most 10 relevant text snippets $s_{i,1}$, $s_{i,2}$, $s_{i,3}$,... from the returned articles. Again, the list should be ordered by decreasing confidence. A single snippet list will be returned per question and participating system, and the list may contain any number (or no) snippets from any of the returned articles $d_{i,1}$, $d_{i,2}$, $d_{i,3}$,... Each snippet will be represented by the unique identifier of the article it comes from, the identifier of the section the snippet starts in, the offset of the first character of the snippet in the section the snippet starts in, the identifier of the section the snippet ends in, and the offset of the last character of the snippet in the section the snippet ends in. The snippets themselves will also have to be returned (as strings).
    \end{itemize}

    \textbf{Phase A+}. In Phase A+ of Task 13b, the participants will be provided with English questions as in Phase A (above), but will be required to return Exact and/or Ideal answers, as in for Phase B (below).

    \textbf{Phase B}. The participants will be provided with the same questions $q_1$, $q_2$,...,$q_n$ as in Phase A, but this time they will also be given gold (correct) lists of articles and snippets. The "gold" lists will contain articles and snippets identified by biomedical experts as relevant and providing enough information to answer the questions. For each question, each participating system may return an ideal answer, i.e., a paragraph-sized summary of relevant information. In the case of yes/no, factoid, and list questions, the systems may also return exact answers; for summary questions, no exact answers will be returned. The participants will be told the type of each question. A participating system may return only Exact answers, or only Ideal answers, or (ideally) both Exact and Ideal answers. Definition of Exact answers and Ideal answers can be found in Section~\ref{sec:data}.

    \textbf{Objective}. Systems may participate in any of the three phases mentioned above. In this report, we mainly focus on Phase A and Phase A+, and our goal is to develop a system for BioASQ 13B capable of:
    \begin{itemize}
        \item Accurately retrieving relevant documents and snippets (Phase A).
        \item Providing high-quality Exact and Ideal answers for biomedical questions (Phase A+).
        \item Providing results of Phase A and Phase A+ within 1 day after the test set is released.
    \end{itemize}

\section{Data Description}
    \label{sec:data}

    Task 13B uses benchmark datasets containing training and test questions, in English, along with gold standard (reference) answers. The benchmark datasets are being constructed by a team of biomedical experts from around Europe. In this report, we download historical datasets of BioASQ, ranging from Task 2B to Task 13B. Please note that the test data for the 13B model has been released in batches on a bi-weekly basis starting from March 26, 2025. As of now (June 26, 2025), no ground truth labels have been provided.

    \begin{table*}[th]
        \footnotesize
        \caption{Number of unique questions and articles in the dataset.}
        \label{tab:dataset}
        \centering
        \setlength{\tabcolsep}{4mm}{
        \begin{tabular}{ccccccc}
            \toprule
            \multicolumn{1}{c}{Edition} & \multicolumn{3}{c}{Questions} & \multicolumn{3}{c}{Articles}\\
                    \cmidrule(lr){2-4} \cmidrule(lr){5-7}  ~ & Training & Test & Shared & Training & Test & Shared\\
            \midrule
            Task 2B & 306 & 500 & 0 & 4194 & 5948 & 49\\
            Task 3B & 806 & 497 & 0 & 10093 & 5926 & 146\\
            Task 4B & 1303 & 496 & 0 & 15873 & 4174 & 112\\
            Task 5B & 1799 & 500 & 0 & 19935 & 5798 & 160\\
            Task 6B & 2251 & 500 & 0 & 25573 & 3167 & 86\\
            Task 7B & 2747 & 500 & 0 & 28968 & 2278 & 93\\
            Task 8B & 3243 & 500 & 0 & 31153 & 2298 & 123\\
            Task 9B & 3743 & 497 & 0 & 33330 & 3523 & 154\\
            Task 10B & 4234 & 486 & 0 & 36844 & 3478 & 127\\
            Task 11B & 4719 & 330 & 0 & 40221 & 2844 & 54\\
            Task 12B & 5049 & 340 & 0 & 43169 & 6570 & 181\\
            Task 13B & 5389 & 340 & 0 & 49610 & - & -\\
            \bottomrule
        \end{tabular}}
    \end{table*}

    Table~\ref{tab:dataset} shows the number of unique questions and articles in the datasets, where "Shared" means questions or articles that appear in both training and test sets of the corresponding edition. We observe that each year's training set is a combination of the previous year's training and test sets. Therefore, \textbf{we focus on the datasets in Task 12B}, which already include all available questions.

    \begin{figure}[!th]
        \centering
        \includegraphics[width=0.9\linewidth]{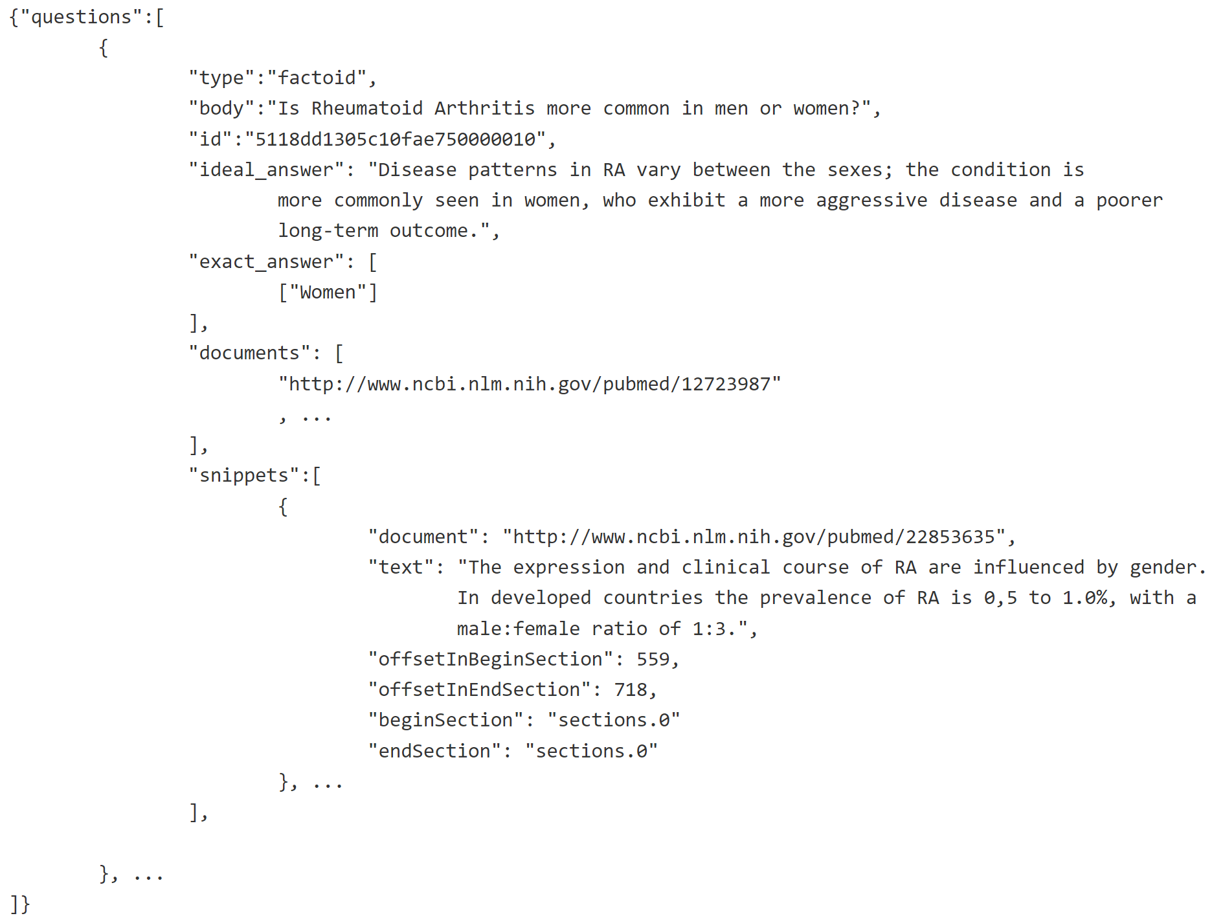}
        \vspace{-5mm}
        \caption{Data sample in JSON format.}\label{fig:data}
    \end{figure}

    \textbf{Types of Questions}. The benchmark datasets contain four types of questions:
    \begin{itemize}
        \item \textbf{Yes/No} questions: These are questions that, strictly speaking, require "yes" or "no" answers, though of course in practice longer answers will often be desirable. For example, "Do CpG islands colocalise with transcription start sites?" is a yes/no question.
        \item \textbf{Factoid} questions: These are questions that, strictly speaking, require a particular entity name (e.g., of a disease, drug, or gene), a number, or a similar short expression as an answer, though again a longer answer may be desirable in practice. For example, "Which virus is best known as the cause of infectious mononucleosis?" is a factoid question.
        \item \textbf{List} questions: These are questions that, strictly speaking, require a list of entity names (e.g., a list of gene names), numbers, or similar short expressions as an answer; again, in practice additional information may be desirable. For example, "Which are the Raf kinase inhibitors?" is a list question.
        \item \textbf{Summary} questions: These are questions that do not belong in any of the previous categories and can only be answered by producing a short text summarizing the most prominent relevant information. For example, "What is the treatment of infectious mononucleosis?" is a summary question.
    \end{itemize}

    \textbf{Types of Answers}. The benchmark datasets contain two types of answers: Exact answers and Ideal answers. The Exact answer consists of short phrases and varies depending on the type of question:
    \begin{itemize}
        \item For each Yes/No question, the exact answer of each participating system will have to be either "yes" or "no".
        \item For each Factoid question, each participating system will have to return a list* of up to 5 entity names (e.g., up to 5 names of drugs), numbers, or similar short expressions, ordered by decreasing confidence.
        \item For each List question, each participating system will have to return a single list* of entity names, numbers, or similar short expressions, jointly taken to constitute a single answer (e.g., the most common symptoms of a disease). The returned list will have to contain no more than 100 entries of no more than 100 characters each.
        \item No exact answers will be returned for summary questions.
    \end{itemize}
    The Ideal answer is a single paragraph-sized text ideally summarizing the most relevant information from articles and snippets. For each question (yes/no, factoid, list, summary), each participating system of Phase B may return an ideal answer. Each returned "ideal" answer is intended to approximate a short text that a biomedical expert would write to answer the corresponding question (e.g., including prominent supportive information), whereas the "exact" answers are only "yes"/"no" responses, entity names or similar short expressions, or lists of entity names and similar short expressions; and there are no "exact" answers in the case of summary questions. The maximum allowed length of each "ideal" answer is 200 words.

    \textbf{Data Structure}. The training and test datasets use the JSON format as shown in Figure~\ref{fig:data}, where "type" denotes type of the question, "body" denotes the question, "id" is the unique identifier of this question, "ideal\_answer" is a list of the gold standard Ideal answers, "exact\_answer" is the gold standard Exact answer, "documents" is a list of gold standard articles from PubMed that closely related to the question, and "snippets" is a list of sentences that related to the question from the gold standard article abstracts. 

    In the case of Yes/No questions, the "exact\_answer" field is either yes or no, with no explanations.

    In the case of Factoid questions, the "exact\_answer" field is a list of lists. Each of the inner list (up to 5 inner lists are allowed) should contain the name of the entity (or number, or other similar short expression) sought by the question; Since BioASQ5, no multiple names (synonyms) should be submitted for any entity, therefore each inner list should only contain one element. If the List contains more than one elements, only the first element will be taken into account for evaluation. Note that in the training data, the field is a \*simple\* list containing the entity (and its synonyms if they were identified) that answers the question.

    In the case of List questions, similar to Factoid questions, the "exact\_answer" field is a list of lists. Each element of the outermost list is a list corresponding to one of the entities (or numbers, or other similar short expressions) seeked by the question. Since BioASQ5, no multiple names (synonyms) should be submitted for any entity, therefore, each inner list should only contain one element. If the List contains more than one elements, only the first element will be taken into account for evaluation.

\section{Methodology}
    \label{sec:method}

    In this section, we first introduce the overall framework of the proposed method, then go into details of each part.

    \subsection{Overall Framework}
    
    The proposed method is structured as a multi-stage framework that systematically addresses the challenge of biomedical question answering. At its core, the framework integrates LLM-driven keyword extraction, advanced document filtering, knowledge graph augmentation, and adaptive answering strategies, ensuring that each stage contributes to the precision and relevance of the final answer.

    The process begins by extracting keywords from both the input questions and the answers generated by a large language model. This initial stage ensures that essential concepts and entities from both the user’s query and the LLM’s interpretations are represented. These keywords are then combined in various ways to form robust search queries, which are used to retrieve potentially relevant documents from large biomedical databases.
    
    Once a diverse set of documents is collected, the next phase involves reranking the results to prioritize those most likely to contain valuable information. This reranking step is further refined by applying a language model-based filtering process, which evaluates the content and relevance of the highest-ranked documents and retains only those most likely to support effective question answering. In parallel, the framework leverages a knowledge graph to augment the document set, identifying additional sources that may be closely related to the question but not captured through keyword-based retrieval alone.
    
    With a comprehensive pool of relevant documents in place, the framework then adapts its answering strategy based on the specific type of question being posed. By tailoring the approach for Yes/No, Factoid, List, and Summary questions, the system maximizes the accuracy and specificity of its responses. Each of these components is described in greater detail in the following subsections, which elaborate on their individual contributions and integration within the overall pipeline.

    \subsection{Keyword Extraction}
    \label{sec:keyword}

    Keyword extraction serves as the foundational step in our framework, playing a crucial role in enabling effective document retrieval in subsequent stages. Our approach extracts essential keywords from both the input question and the initial answer produced by a large language model (LLM). By capturing keywords from these two complementary sources, we maximize the chances of retrieving all relevant documents that contain the necessary information.

    \textbf{Keywords in Questions}. For each question, we employ a single LLM prompt to extract multiple types of keyword information. First, we identify all keywords within the question to capture the full breadth of potentially important concepts. Since not every term is equally central to the query’s meaning, we also determine the one or two keywords that are most significant. This targeted identification allows for more focused retrieval when a narrower or more precise search is required. Furthermore, we single out the most important keyword from the question, which is especially useful for tasks that demand high precision.
    
    \textbf{Keywords in Answers}. In addition to extracting keywords from questions, we also process the LLM-generated answers to identify all relevant keywords they contain. By integrating keywords found in the answers, our method captures critical details that may have been inferred or clarified, thus broadening the scope and depth of the document retrieval process.
    
    \textbf{Synonyms of Keywords}. Another important consideration is the presence of synonyms or alternate terms for key entities, which are frequently used in literature instead of the exact wording found in questions. To ensure comprehensive retrieval, we apply two strategies to expand our keyword sets with synonyms. The first strategy relies on a curated synonym database to generate alternative expressions for the extracted keywords. The second strategy leverages the LLM’s ability to suggest additional synonyms, tapping into its semantic understanding to uncover related terms that may fall outside our predefined database. These complementary approaches significantly improve recall by addressing terminological variation.
    
    Through this combination of thorough keyword extraction and strategic synonym expansion, our framework establishes a robust basis for accurate and wide-ranging document retrieval, effectively supporting the later components of our system.

    \subsection{Document Retrieval}
    \label{sec:retrieval}

    Building on the keyword extraction process described in Section~\ref{sec:keyword}, we employ the \textbf{iSearch}\footnote{\href{https://biokde.insilicom.com/}{https://biokde.insilicom.com/}} system to retrieve documents relevant to each question from the PubMed database. Each retrieved document includes key bibliographic metadata such as the PMID, publication year, title, and abstract, providing a rich context for subsequent answer generation.

    To ensure broad and accurate coverage, we design nine distinct retrieval strategies based on different combinations of extracted keywords. The first seven strategies are grounded in variations of keywords from the question itself:
    \begin{itemize}
        \item q: all keywords in the question.
        \item iq: the one or two most important keywords.
        \item i1q: the single most important keyword.
        \item raw\_q: the raw question text.
        \item syn: keywords and their synonyms from the synonym database.
        \item syn\_llm: keywords and their synonyms identified by the LLM.
        \item isyn\_llm: most important keyword and its synonym based on LLM’s results.
    \end{itemize}
    
    The remaining two strategies incorporate answer-derived keywords:
    \begin{itemize}
        \item q1a: for each answer entity, we retrieve documents using the combined keywords from the question and that entity, aggregating all retrieved documents.
        \item qa: we use the combined set of keywords from both the question and all answer entities in a single query.
    \end{itemize}
    
    After performing searches using all eight strategies, we select the top ranked documents from each approach based on relevance. The results from all strategies are then aggregated to form the final document pool. This comprehensive multi-strategy retrieval process ensures that the system captures a wide spectrum of relevant literature, increasing the likelihood of retrieving articles that contain the necessary information, regardless of how entities are mentioned or referenced in the source texts. By integrating question-based and answer-based perspectives, our document retrieval component establishes a strong and flexible foundation for effective information extraction and answer synthesis in downstream tasks.

    \subsection{Document Reranking and Filtering}
    \label{sec:rerank}

    To enhance the accuracy of document selection while maintaining computational efficiency, we implement a multi-stage reranking and filtering process on the documents retrieved as described in Section 3.3. The goal of this approach is to systematically prioritize the most relevant documents for question answering, ensuring that only high-quality, informative sources are passed on to subsequent stages.

    \textbf{Reranking}. In the first stage, we utilize \textit{jina-reranker-v2-base-multilingual}\footnote{\href{https://huggingface.co/jinaai/jina-reranker-v2-base-multilingual}{https://huggingface.co/jinaai/jina-reranker-v2-base-multilingual}}~\cite{sturua2024jina} to rerank the initial pool of retrieved documents. \textit{jina-reranker-v2-base-multilingual} is a transformer-based cross-encoder that has been fine-tuned specifically for text reranking tasks. The model processes a query and a document pair—where the document consists of the concatenated title and abstract—and outputs a score reflecting the document’s relevance to the query. Due to its training on large-scale, multilingual datasets of query-document pairs, \textit{jina-reranker-v2-base-multilingual} is highly effective in sorting biomedical documents by their potential utility for answering the input question.
    
    \textbf{Reranking Again}. To further refine the ranking, we apply \textit{bge-reranker-v2.5-gemma2-lightweight}\footnote{\href{https://huggingface.co/BAAI/bge-reranker-v2.5-gemma2-lightweight}{https://huggingface.co/BAAI/bge-reranker-v2.5-gemma2-lightweight}}~\cite{li2023making,chen2024bge} to the top ranked documents from the first reranking. This advanced multilingual model, derived from \textit{gemma2-9b}, incorporates token compression and layerwise reduction, allowing it to maintain strong performance with improved computational efficiency. Although it demonstrates slightly better accuracy than \textit{jina-reranker-v2-base-multilingual}, the  \textit{bge-reranker-v2.5-gemma2-lightweight} model is more resource-intensive and slower, which restricts its use to only the highest-ranked subset of documents. This staged application of reranking ensures that the system remains within the time constraints required by the BioASQ challenge, while still benefiting from the strengths of both models.
    
    \textbf{LLM Filtering}. Once the documents have been reranked, we employ a large language model (LLM) to filter out irrelevant documents based on their content and relevance to the specific question type. Each document is scored on a scale from 0 to 2, with a score of 2 assigned to documents containing explicit or directly inferable information that answers the question—including mention of key terms or entities. A score of 1 is given to documents that offer contextual or partial information that could reasonably be inferred as relevant, even if not explicitly stated. Documents that do not relate to the question at all are scored as 0. For Yes/No, Factoid, and Summary questions, only documents scored as 2 are retained, except when fewer than ten documents meet this criterion—in which case, those scored as 1 may also be considered. For List questions, both scores of 1 and 2 are accepted, as these often require aggregation of information across multiple relevant sources.
    
    Through this rigorous reranking and filtering pipeline, the system effectively narrows down the document set to those most likely to contribute meaningful and accurate answers, balancing both computational cost and answer quality.

    \subsection{Knowledge Graph Augmentation}
    \label{sec:kg}
    List-type questions often pose a significant challenge where answers require multiple specific entities, usually drugs, chemicals, etc. To address this problem, we introduce knowledge graph augmentation. Insilicom's knowledge graph~\cite{zhang2025ikraph} is designed to store the relations between pairs of entities, for example, searching for arsenic and lung cancer will return documents that contain information containing the two. This type of structured data is suitable for answering biomedical questions that present a specific entity and require the return of multiple other entities, each bearing a relation to the proposed entity. To take advantage of the knowledge graph, we first ask an LLM to generate KG queries from a set of questions. The KG queries must then be executed to retrieve documents containing the desired answers.
    
    \textbf{Generating Queries}. For this step, we use Anthropic's \textit{claude-3-7-sonnet-20250219} to analyze the original question and generate a KG-compatible query that identifies the following:
    \begin{itemize}
        \item Entity 1 name: Specifies the name of the first entity in the relation.
        \item Entity 1 type: Specifies the first entity's type
        \item Entity 2 type: Specifies the second entity's type
        \item Entity 2 name: The name of the second entity is optional as questions often ask for more than one entity that belong to the same type (Entity 2 type)
        \item Relation type: Does one entity inhibit or increase the expression of the other or neither? (e.g. positive, negative, association)
        \item Relation direction: which entity affects the other (e.g. 1->2 or 2->1)
    \end{itemize}
    This information helps narrow the search space to the appropriate scope and avoid inclusion of inaccurate answers.
    
    \textbf{Querying Knowledge Graph}. Because the KG data are primarily stored in a Mongo database, it is necessary to convert the queries from the previous step into Mongo queries before executing. Once converted and executed, the returned documents can be further filtered and processed as needed.

    This two-step process comprises the knowledge graph augmentation method, which allows the pipeline to leverage the power of Insilicom's preexisting biomedical KG to provide supporting documents that contain favorable answers not easily extracted from other sources.

    \subsection{Question Answering}
    \label{sec:qa}

    To fully leverage the set of relevant documents identified in Sections 3.4 and 3.5, we design a comprehensive question answering pipeline that combines evidence extraction and tailored answering strategies for each question type. The process begins by using a large language model (LLM) to extract the most pertinent pieces of evidence from each retrieved document, focusing specifically on passages or details that directly address the target question. This evidence-driven approach ensures that subsequent reasoning is firmly grounded in the supporting literature.

    \textbf{Yes/No Questions}. We organize the extracted evidence by grouping all supporting documents into two clusters: those that indicate a “yes” answer and those that indicate a “no” answer. The LLM then assesses the reasoning and strength of evidence for each side, ultimately determining which answer is more substantiated by the literature. In situations where the question pertains to rapidly evolving topics or issues sensitive to new developments, the publication years of the supporting documents are explicitly considered to prioritize more recent, potentially more accurate, sources.
    
    \textbf{Factoid Questions} require a different treatment. After extracting evidence, we group the documents into clusters, each supporting a distinct factoid relevant to the question. The LLM evaluates and ranks these clusters based on the relevance and quality of the supporting evidence. When the relevance of certain factoids is ambiguous or disputed, we further consult the publication dates, favoring information from newer or more authoritative sources to resolve uncertainties and improve answer reliability.
    
    \textbf{List Questions}. The pipeline focuses on entity extraction. \textbf{\textit{Step 1}}: The LLM identifies and extracts lists of relevant entities from each document based on the provided evidence. \textbf{\textit{Step 2}}: As the process continues, these entities are merged into a global candidate list, with synonyms incorporated into corresponding sublists to account for terminological variation. \textbf{\textit{Step 3}}: In the final step, the LLM reviews all entities in the merged list, selecting only a single representative entity from each group of synonyms and retaining those that constitute valid, specific answers to the biomedical question at hand.
    
    \textbf{Summary Questions} are handled through a process of synthesis. Evidence extracted from the supporting documents is grouped into clusters, each of which supports a similar line of reasoning or answer. The LLM then generates concise summaries for each group, capturing the core insights or findings. Finally, these intermediate summaries are synthesized into a single, comprehensive answer that directly addresses the original question, ensuring that the response is both thorough and informed by the breadth of available evidence.
    
    Through this adaptive, evidence-based methodology, the question answering component provides robust and context-sensitive answers across a diverse set of biomedical question types, ultimately enhancing the quality and credibility of the system’s responses.

    \subsection{Speed Optimization}
    
    Efficient processing is critical for deployment of question answering systems in BioASQ, especially when working with computationally intensive models such as LLMs and advanced rerankers. To address potential bottlenecks and accelerate the overall workflow, we implement several speed optimization strategies within our framework.

    A key approach to reducing runtime is the adoption of parallel processing for the language model. In both the document filtering and question answering stages, we leverage the OpenAI Asyncio API\footnote{\href{https://platform.openai.com/docs/api-reference/batch}{https://platform.openai.com/docs/api-reference/batch}}, which enables asynchronous execution of LLM queries. By distributing the processing of multiple documents and questions across concurrent tasks, we are able to significantly decrease the wall-clock time required for these stages, even when handling large batches of data.
    
    In addition to parallelizing language model operations, we further optimize efficiency by running the \textit{bge-reranker-v2.5-gemma2-lightweight} model and the question answering module concurrently. Rather than executing these steps sequentially, our system initiates the reranking of documents while simultaneously starting the question answering process on the outputs from previous steps. This concurrent execution model ensures that system resources are fully utilized and that no component is left idle while waiting for another to complete.
    
    Collectively, these speed optimization measures enable our framework to process large-scale biomedical question answering tasks with substantially improved throughput. As a result, we are able to meet the stringent time constraints of BioASQ without sacrificing the depth or accuracy of our results.

\section{Experiments}
    In this section, we first introduce the experimental setting, then we evaluate our proposed method.

    \subsection{Experimental Settings}

    The majority of our proposed method is underpinned by the use of large language models, with GPT-4o~\cite{hurst2024gpt} serving as the backbone throughout all main components. To ensure the stability and reproducibility of results, we set the temperature parameter of the language model to zero. This deterministic configuration minimizes randomness in the model’s outputs, which is particularly important for a fair comparison across experimental runs.

    For the document retrieval stage, we limit our search to documents published in 1990 or later, utilizing the iSearch system as our retrieval engine. To balance retrieval speed and comprehensiveness, we focus on six specific retrieval strategies: `q', `iq', `i1q', `raw\_q', `isyn\_llm', and `q1a'. For each strategy, we collect the top 5000 documents, ensuring a sufficiently broad yet manageable pool for subsequent reranking and filtering steps.
    
    In the document reranking phase, all retrieved documents are first processed by the \textit{jina-reranker-v2-base-multilingual} model using its default hyperparameter settings. The top 1000 documents from this initial reranking are then further refined with the \textit{bge-reranker-v2.5-gemma2-lightweight} model, which is configured with cutoff\_layers set to [28], compress\_ratio to 2, and compress\_layer to [24, 40]. These parameters were selected to strike a balance between computational efficiency and ranking quality.
    
    To determine which documents advance to the LLM filtering stage, we introduce dual cutoff criteria based on both the ranking scores and absolute rankings produced by \textit{jina-reranker-v2-base-multilingual}. For Yes/No questions, documents are included if they have a ranking score above 0.5 or appear in the top 1000. For Factoid questions, the thresholds are a score above 0.5 or inclusion in the top 1500. For List questions, more stringent cutoffs are used, with a required score above 0.6 or a place in the top 1500. Summary questions adopt a similar approach, retaining documents with a score above 0.6 or a top 1000 ranking. Documents meeting either of these thresholds are selected for further evaluation by the language model, allowing for flexible yet consistent filtering across diverse question types.

    \subsection{Comparison Methods}

    To comprehensively evaluate the performance of our proposed framework, we design five distinct comparison methods, corresponding to the five allowable submissions in the BioASQ challenge. Each method represents a different combination of document retrieval, reranking, and knowledge integration, allowing us to assess the impact of each component on overall question answering quality.

    \textbf{Naive}. The first comparison method, referred to as Naive, relies solely on the capabilities of the large language model to answer questions directly, without leveraging any supporting documents. For Phase A, however, we still retrieve the top 10 documents using the \textit{jina-reranker-v2-base-multilingual} as described in Section~\ref{sec:rerank}, even though these documents are not provided to the language model for answer generation.
    
    \textbf{Baseline\_Top10} constitutes our second comparison method. Here, for each question, we supply the large language model with the top 10 documents selected by \textit{bge-reranker-v2.5-gemma2-lightweight}, as detailed in Section~\ref{sec:rerank}. The model is then tasked with answering the question using only the information present in the given set of documents.
    
    \textbf{Baseline\_Top20} increases the document set to the top 20 retrieved by \textit{bge-reranker-v2.5-gemma2-lightweight} for each question. The language model uses this broader evidence base to generate its answers. For consistency in Phase A, however, we still use only the top 10 documents for retrieval and analysis.
    
    \textbf{Mainpipeline} represents our full retrieval and answering pipeline as described in Section~\ref{sec:method}, but excludes the knowledge graph augmentation outlined in Section~\ref{sec:kg}. In this approach, document selection is adapted to the type of question. For Yes/No and Factoid questions, the top 10 documents within the cluster identified by the language model are chosen. For List questions, we select documents whose evidential content fully covers the set of final answers, while for Summary questions, we focus on the top 10 documents that provide the strongest support for the synthesized response.
    
    \textbf{KG} method extends the Mainpipeline by incorporating knowledge graph augmentation specifically for List questions. Since this augmentation does not alter the processing of Yes/No, Factoid, or Summary questions, we introduce an additional evaluation strategy for the KG method. For Yes/No questions, we employ a majority vote among the answers produced by the Naive, Baseline\_Top10, Baseline\_Top20, and Mainpipeline methods, ensuring a diverse and robust assessment. This multi-faceted comparative framework enables a rigorous analysis of each method’s strengths and limitations, highlighting the contributions of both document retrieval strategies and knowledge graph augmentation.

    It is important to emphasize that, as detailed in Section~\ref{sec:qa}, we have systematically extracted supporting evidence from each document identified as relevant by our retrieval and filtering processes. These extracted evidences, which encapsulate the most pertinent information related to the given question, serve a critical role in our framework. Specifically, for Phase A of the evaluation, these evidences will be directly provided as the snippets for each answer.
    
    \subsection{Evaluation Metrics}

    \textbf{Phase A}. The task requires a list of at most 10 relevant articles and a list of at most 10 relevant text snippets as output. Articles are evaluated by \textit{Mean Average Precision (MAP)}:
    \begin{equation}
        AP = \frac{\sum_{r=1}^{|L|} P(r)\cdot rel(r)}{min(|L_R|, 10)},
    \end{equation}
    \begin{equation}
        MAP = \frac{1}{n} \sum_{i=1}^{n} AP_i,
    \end{equation}
    where $|L|$ is the number of predicted articles in the list, $|L_R|$ is the number of relevant articles, $P(r)$ is the precision when the returned list is treated as containing only its first $r$ items, and $rel(r)$ equals 1 if the $r$-th article of the list is in the golden set (i.e., if the $r$-th item is relevant) and 0 otherwise.

    For snippets, the evaluation metric is \textit{F-measure}. A snippet is determined by the article it comes from and by the offsets (positions) in the article of the first and last characters of the snippet. Let $S$ be the set of all the article-offset pairs of all the characters in the snippets returned by a system for a particular question, $G$ the set of all the article-offset pairs of all the characters in the golden snippets of the question, and let $|s|$ denote the cardinality of a set $s$. The definitions of precision $P_{snip}$ and recall $R_{snip}$ for snippets are:
    \begin{equation}
        P_{snip} = \frac{|S \cap G|}{|S|},
    \end{equation}
    \begin{equation}
        R_{snip} = \frac{|S \cap G|}{|G|}.
    \end{equation}
    In effect, $P_{snip}$ divides the size (in characters) of the total overlap between the returned and golden snippets by the total size of the returned snippets, whereas $R_{snip}$ divides the size of the total overlap by the total size of the golden snippets. The definitions of mean precision, mean recall, and mean \textit{F-measure} for snippets are the same as the traditional definitions, but they use $P_{snip}$ and $R_{snip}$ instead of precision and recall.

    \textbf{Phase A+}. The task requires Exact or Ideal answers as output. Different types of questions are evaluated by different measures, stated as follows.

    For yes/no questions, the evaluation metric is \textit{Macro F1}, the unweighted average of the F1 scores computed separately for the \textit{yes} and \textit{no} classes:
    \begin{equation}
        F1_{macro} = \frac{F1_{yes} + F1_{no}}{2},
    \end{equation}
    where $F1_{yes}$ and $F1_{no}$ are each the harmonic mean of the precision and recall of the corresponding class. By weighting both classes equally regardless of how frequently they occur, \textit{Macro F1} gives a more balanced assessment than raw accuracy when the yes/no labels are unevenly distributed.

    For Factoid questions, the evaluation metric is \textit{Mean Reciprocal Rank (MRR)}:
    \begin{equation}
        MRR = \frac{1}{n}\sum_{i=1}^{n} \frac{1}{r(i)}.
    \end{equation}
    In the definition above, for each factoid question $q_i$, we search the returned list looking for the topmost position that contains the golden entity name (or one of its synonyms) based on exact matching. If the topmost position is the $j$-th one, then $r(i)=j$; otherwise $r(i) \to +\infty$, i.e., $\frac{1}{r(i)}=0$. In effect, \textit{MRR} rewards systems that manage to include the golden responses (or their synonyms) higher in the returned lists.

    For List questions, the evaluation metric is \textit{F-measure}, where truth positives are calculated based on exact matching.

    For Summary questions, the initial score is calculated by \textit{ROUGE-2} and \textit{ROUGE-SU4} using word $n$-grams when computing the overlap between an automatically constructed summary $S$ and a set $Refs$ of reference summaries:
    \begin{equation}
        ROUGE-N(S|Refs) = \frac{\sum_{R \in Refs} \sum_{g_n \in R} C(g_n,S,R)}{\sum_{R \in Refs} \sum_{g_n \in R} C(g_n,R)}.
    \end{equation}
    In the definition above, $g_n$ is a word $n$-gram, $C(g_n,S,R)$ is the number of times that $g_n$ co-occurs in $S$ and a reference summary $R$, and $C(g_n,R)$ is the number of times $g_n$ occurs in reference $R$. \textit{ROUGE-2} is\textit{ ROUGE-N} with $n=2$; and \textit{ROUGE-SU4} is a version of \textit{ROUGE-SU} with the maximum distance between the words of any skip bigram limited to 4.

    \textbf{Final Evaluation}. The BioASQ committee recruits experts to manually review the submitted results. They adjust the scores for Yes/No, Factoid, and List questions, and assess the answers to Summary questions based on the following four criteria:
    \begin{itemize}
        \item Information recall: whether all necessary information is reported;
        \item Information precision: whether no irrelevant information is reported;
        \item Information repetition: whether the answer does not repeat the same information multiple times;
        \item Readability: whether the answer is easily readable and fluent.
    \end{itemize}
    Each criterion is scored on a scale from 1 to 5, and the final score is calculated as the average of these four individual scores.

    \subsection{Results and Analysis}

    \begin{table*}[ht]
        \footnotesize
        \caption{Results for BioASQ Task 13B Phase A document retrieval.}
        \label{tab:PhaseA_Doc}
        \centering
        \setlength{\tabcolsep}{2mm}{
        \begin{tabular}{c|ccccc|c}
            \toprule
            \multicolumn{1}{c|}{Phase A} & \multicolumn{5}{c}{\textit{Mean Average Precision}} & \multicolumn{1}{|c}{Our}\\
                    \cmidrule(lr){2-6}  Document & Naive & Baseline\_Top10 & Baseline\_Top20 & Mainpipeline & KG & Rank\\
            \midrule
            Batch 1 & 0.5219 & \textbf{0.5627} & \textbf{0.5627} & 0.4796 & 0.4800 & 1\\
            Batch 2 & 0.5945 & \textbf{0.6768} & \textbf{0.6768} & 0.5423 & 0.5299 & 1\\
            Batch 3 & 0.5193 & \textbf{0.5757} & \textbf{0.5757} & 0.4513 & 0.4571 & 1\\
            Batch 4 & 0.3930 & \textbf{0.4847} & \textbf{0.4847} & 0.4051 & 0.3990 & 1\\
            \bottomrule
        \end{tabular}}
    \end{table*}
    \begin{table*}[ht]
        \footnotesize
        \caption{Results for BioASQ Task 13B Phase A snippet extraction.}
        \label{tab:PhaseA_Snp}
        \centering
        \setlength{\tabcolsep}{2mm}{
        \begin{tabular}{c|ccccc|c}
            \toprule
            \multicolumn{1}{c|}{Phase A} & \multicolumn{5}{c}{\textit{F-Measure}} & \multicolumn{1}{|c}{Our}\\
                    \cmidrule(lr){2-6}  Document & Naive & Baseline\_Top10 & Baseline\_Top20 & Mainpipeline & KG & Rank\\
            \midrule
            Batch 1 & 0.2140 & 0.2314 & 0.2314 & \textbf{0.2366} & 0.2270 & 1\\
            Batch 2 & 0.2371 & 0.2729 & 0.2729 & \textbf{0.2746} & 0.2684 & 1\\
            Batch 3 & 0.2218 & \textbf{0.2501} & \textbf{0.2501} & 0.2319 & 0.2355 & 1\\
            Batch 4 & 0.1942 & \textbf{0.2164} & \textbf{0.2164} & 0.1926 & 0.1916 & 1\\
            \bottomrule
        \end{tabular}}
    \end{table*}
    \begin{table*}[ht]
        \footnotesize
        \caption{Results for BioASQ Task 13B Phase A+ Yes/No questions.}
        \label{tab:PhaseA+_yesno}
        \centering
        \setlength{\tabcolsep}{2mm}{
        \begin{tabular}{c|ccccc|c}
            \toprule
            \multicolumn{1}{c|}{Phase A} & \multicolumn{5}{c}{\textit{Macro F1}} & \multicolumn{1}{|c}{Our}\\
                    \cmidrule(lr){2-6}  Document & Naive & Baseline\_Top10 & Baseline\_Top20 & Mainpipeline & KG & Rank\\
            \midrule
            Batch 1 & \textbf{1.0000} & 0.9244 & 0.9244 & 0.9244 & 0.9244 & 1\\
            Batch 2 & 0.5882 & \textbf{0.8712} & \textbf{0.8712} & \textbf{0.8712} & \textbf{0.8712} & >2\\
            Batch 3 & 0.7270 & 0.7708 & \textbf{0.8362} & 0.7708 & 0.7708 & >2\\
            Batch 4 & 0.8756 & \textbf{0.9563} & \textbf{0.9563} & 0.9097 & \textbf{0.9563} & 2\\
            \bottomrule
        \end{tabular}}
    \end{table*}
    \begin{table*}[ht]
        \footnotesize
        \caption{Results for BioASQ Task 13B Phase A+ Factoid questions.}
        \label{tab:PhaseA+_factoid}
        \centering
        \setlength{\tabcolsep}{2mm}{
        \begin{tabular}{c|ccccc|c}
            \toprule
            \multicolumn{1}{c|}{Phase A} & \multicolumn{5}{c}{\textit{Mean Reciprocal Rank}} & \multicolumn{1}{|c}{Our}\\
                    \cmidrule(lr){2-6}  Document & Naive & Baseline\_Top10 & Baseline\_Top20 & Mainpipeline & KG & Rank\\
            \midrule
            Batch 1 & 0.1827 & 0.3269 & 0.3654 & \textbf{0.3782} & \textbf{0.3782} & >2\\
            Batch 2 & 0.3333 & 0.4506 & \textbf{0.5185} & 0.4012 & 0.4012 & >2\\
            Batch 3 & 0.3500 & 0.2500 & 0.2750 & \textbf{0.3667} & \textbf{0.3667} & >2\\
            Batch 4 & 0.4091 & \textbf{0.5000} & 0.4318 & 0.4356 & 0.4356 & >2\\
            \bottomrule
        \end{tabular}}
    \end{table*}
    \begin{table*}[ht]
        \footnotesize
        \caption{Results for BioASQ Task 13B Phase A+ List questions.}
        \label{tab:PhaseA+_list}
        \centering
        \setlength{\tabcolsep}{2mm}{
        \begin{tabular}{c|ccccc|cc}
            \toprule
            \multicolumn{1}{c|}{Phase A} & \multicolumn{5}{c}{\textit{F-measure}} & \multicolumn{1}{|c}{Recall} & \multicolumn{1}{c}{F-measure}\\
                    \cmidrule(lr){2-6}  Document & Naive & Baseline\_Top10 & Baseline\_Top20 & Mainpipeline & KG & Rank & Rank\\
            \midrule
            Batch 1 & 0.2496 & \textbf{0.3038} & 0.2782 & 0.1698 & 0.1645 & >2 & >2\\
            Batch 2 & 0.2828 & \textbf{0.3758} & 0.3647 & 0.2829 & 0.2796 & 1 & 2\\
            Batch 3 & 0.2344 & 0.4238 & \textbf{0.4259} & 0.3195 & 0.3642 & 1 & >2\\
            Batch 4 & 0.2837 & 0.3307 & \textbf{0.3578} & 0.2703 & 0.2533 & 1 & 1\\
            \bottomrule
        \end{tabular}}
    \end{table*}

    The evaluation of our proposed framework was conducted using the official BioASQ test sets, which are released in four separate batches on a bi-weekly schedule beginning March 26, 2025. It is important to note that, as of July 2025, only machine-evaluated results have been made publicly available. Consequently, the following analysis is based on the currently accessible results and may be subject to revision once the final, human-assessed scores are released. The results for the Summary questions in Phase A+ are not included, as they heavily rely on human evaluation. Since BioASQ awards only the top two teams in each task, we focus our analysis on the relative ranking of our submissions rather than absolute metric values alone.
    
    Tables~\ref{tab:PhaseA_Doc} and \ref{tab:PhaseA_Snp} summarize the results obtained for Phase A of the competition, which focus on document retrieval and snippet extraction tasks, respectively. Our system delivered the strongest performance among all participating teams across both tasks, ranking first in every one of the four batches for document retrieval as well as for snippet extraction. For document retrieval, measured by \textit{Mean Average Precision (MAP)}, our best configuration consistently surpassed all competing teams. For snippet extraction, evaluated by \textit{F-measure}, our methods again topped every batch, with the Mainpipeline variant leading in Batches 1 and 2 and the Baseline variants leading in Batches 3 and 4. These outcomes suggest that our document retrieval and reranking strategy has been instrumental in enhancing the relevance and quality of the retrieved documents.
    
    For Phase A+, the results are presented in Tables~\ref{tab:PhaseA+_yesno}, \ref{tab:PhaseA+_factoid}, and \ref{tab:PhaseA+_list}. For Yes/No questions, evaluated by \textit{Macro F1}, our framework was highly competitive: our \textit{Naive} (LLM-only) variant attained a perfect score of 1.0000 and ranked first in Batch 1, although performance on Batches 2 and 3 was more middle-of-the-pack. For Factoid questions, measured by \textit{Mean Reciprocal Rank (MRR)}, our best variant placed third at the team level in three of the four batches. For List questions, we report both F-measure ranking and recall ranking, since our system was intentionally designed to favor higher recall in this task. In biomedical list-type QA, missing correct entities can substantially reduce answer completeness, so our retrieval and candidate generation steps prioritize broad coverage of potentially relevant entities. This design led to strong recall performance, with our system ranking first in recall for Batches 2, 3, and 4. Evaluated by \textit{F-measure}, we ranked first in Batch 4 and achieved the highest recall among all teams in Batches 2, 3, and 4. Across Phase A+, the Baseline\_Top10 and Baseline\_Top20 approaches were generally the strongest variants within our suite of methods, especially for List questions. This finding highlights the value of providing a broader set of candidate documents for answer synthesis, suggesting that the inclusion of additional context can be advantageous for complex question types, especially when the document selection and reranking are sufficiently robust. Please note that in \textit{Step 3} of List questions, we initially used a single prompt for entity merging in Batch 1. In Batch 2, this was split into two separate steps: one for selecting synonyms and another for removing irrelevant entities. For Batch 3, we further revised Step 3 by performing the irrelevant entity removal process twice in batches instead of on the whole list of candidates to enhance precision. 
    
    In summary, our framework achieved ten first-place finishes overall---the most of any participating team. Eight of these came in Phase A, where we swept the top rank in all four batches of both the document retrieval and snippet extraction tasks, while the remaining two were obtained in Phase A+ (Yes/No Batch 1, where the perfect score was shared with several teams, and List Batch 4). This comparative analysis highlights the strong competitiveness of our framework. The consistent performance across different batches and tasks demonstrates both the flexibility and the effectiveness of our pipeline, from keyword extraction through to answer generation.

\section{Conclusion}

    The BioASQ challenge continues to drive progress in biomedical semantic indexing and question answering by offering a comprehensive evaluation platform that spans a wide range of research areas, including hierarchical text classification, machine learning, information retrieval, and multi-document summarization. In this technical report prepared for Insilicom LLC, we have addressed the task of biomedical question answering by developing an end-to-end framework that integrates both state-of-the-art language modeling and knowledge graph augmentation.

    Our approach centers around the use of large language models for extracting meaningful keywords from both questions and candidate answers. This dual-source extraction ensures that the document retrieval process is guided by a complete set of relevant terms. We incorporate multiple, complementary retrieval strategies to maximize the diversity and relevance of the document pool. Following retrieval, documents are reranked to prioritize those most likely to contribute to high-quality answers, and the top-ranked candidates are further filtered and evaluated by the language model to assign relevance scores. To further enhance recall and coverage, knowledge graphs are leveraged to supplement the set of relevant documents with additional contextually linked resources.
    
    In the answer synthesis stage, our method leverages the language model not only to extract evidence from the selected documents but also to cluster them according to question type, allowing for targeted aggregation of key information. This clustering-based approach enables the system to deliver answers that are both contextually appropriate and well-supported by the literature. Experimental results highlight the effectiveness and reliability of this pipeline, demonstrating that the combination of LLM-driven document selection, knowledge graph augmentation, and evidence clustering yields robust and accurate biomedical answers.

    Overall, our method achieved ten first-place rankings across all tasks and batches when considering the best-performing run among our submissions. Eight of these first-place rankings were obtained in Phase A document retrieval and snippet extraction, while the remaining two came from Phase A+ Yes/No Batch 1 and List Batch 4. Under the BioASQ award setting, where only the top two teams are recognized, these results demonstrate that our framework was particularly strong in retrieval-oriented tasks and remained competitive in selected answer generation tasks. The methodology presented in this report contributes a novel and practical solution for biomedical QA, illustrating how modern language models and structured knowledge resources can be integrated to address complex information needs in the biomedical domain.

\bibliographystyle{unsrt}  
\bibliography{reference}

\end{document}